# SI-traceable measurement of an optical frequency at low 10$^{-16}$ level without a local primary standard

Hidekazu Hachisu,[1,*] Gérard Petit,[2] Fumimaru Nakagawa,[1] Yuko Hanado[1] and Tetsuya Ido[1]

[1]*Space-Time Standards Laboratory, National Institute of Information and Communications Technology, 4-2-1 Nukui-kitamachi, Koganei, Tokyo, 184-8795, Japan*
[2]*International Bureau of Weights and Measures (BIPM), Pavillon de Breteuil, 92312 Sèvres, France*
*[\*hachisu@nict.go.jp](mailto:hachisu@nict.go.jp)*

**Abstract:** SI-traceable measurements of optical frequencies using International Atomic Time (TAI) do not require a local primary frequency reference, but suffer from an uncertainty in tracing to the SI second. For the measurement of the $^{87}$Sr lattice clock transition, we have reduced this uncertainty to low 10$^{-16}$ level by averaging three sets of ten-day intermittent measurements, in which we operated the lattice clock for 10$^4$ s on each day. Moreover, a combined oscillator of two hydrogen masers was employed as a local flywheel oscillator (LFO) in order to mitigate the impact of sporadic excursion of LFO frequency. The resultant absolute frequency with fractional uncertainty of 4.3×10$^{-16}$ agrees with other measurements based on local state-of-the-art cesium fountains.

# 1. Introduction

Frequency is the most precisely realized quantity in the international system of units (SI) with an accuracy at the $10^{-16}$ level. It is defined on the basis of the $^{133}$Cs microwave transition. For the measurements of optical radiation, femto-second frequency combs accurately convert the optical frequency to a microwave, enabling SI-traceable measurements with the same level of precision. Furthermore, the rapid progress of optical clocks has recently inspired the discussion toward the redefinition of the second using optical transitions [1]. However, it may require nearly a decade to redefine the SI second; until then, we continue to express the optical frequencies with reference to the Cs transition, and such optical frequency measurements will be critically important at the point of the redefinition as the consistency of the SI second should be maintained before and after the transition.

The scale interval of the International Atomic Time (TAI) [2] is a reliable reference to obtain the SI second, although it is a virtual product computed with a latency of 45 days at maximum. So far, nearly ten optical frequency measurements were performed depending on frequency links to TAI [3–7]. The International Bureau of Weights and Measures (BIPM) computes TAI on the basis of the weighted mean of more than 400 atomic clocks worldwide. The correction to the mean free time scale (Échelle Atomique Libre, EAL) is determined with reference to the primary frequency standards (PFS) or the secondary frequency standards, which are operated in national metrological institutes. Thanks to the well-designed method of composing EAL [8], the correction of $6.483\times10^{-13}$ has remained unchanged for more than four years since 2012. The frequency stability of TAI is at the $10^{-16}$ level for an averaging time of more than five days.

It has been recognized that a significant difficulty of TAI-based measurement lies in the requirement of a longer operation time of optical clocks. This is due to several factors. First, the BIPM calculates TAI only at 0:00 UTC on every fifth day, resulting in the TAI frequency being available only as a mean of five days or its integer multiples, so that a similar length of continuous operations is required for the optical clocks. Second, a longer interval of comparison with TAI provides a better uncertainty in the frequency comparison. This is because the satellite-based time comparison between a local clock and TAI has a statistical uncertainty of 0.1 – 0.3 ns depending on the laboratory so that the inferred uncertainty in the frequency comparison is almost inversely proportional to the interval duration [9]. However, this requirement of an interval of at least 5 days may be somewhat relaxed by the intermittent operation of optical clocks distributed over five days with a careful evaluation of the stochastic behavior of local flywheel oscillators (LFO) [10,11]. The LFO is usually a hydrogen-maser (HM) but, in some cases, a cesium fountain standard was used as an LFO, reaching a similar uncertainty [12]. In addition, there is a path to obtaining a calibration of TAI scale interval with respect to the SI second. While the BIPM routinely publishes the calibration of the TAI frequency with an evaluation interval of one month, it is also possible to compute the calibration on a 5-day basis. This technique allowed the total uncertainty of TAI-based measurement to be suppressed below $1\times10^{-15}$ for the first time [13].

In this article, we report the further reduction of the link uncertainty of TAI-based measurement, leading to the absolute frequency of the $^{87}$Sr clock transition with uncertainty below $5\times10^{-16}$. The reduction of the uncertainty was achieved by extending the length of the campaign to 10-days as well as the introduction of a combined LFO. The result of the clock frequency is consistent with other measurements within the uncertainty, indicating a high level of reliability in the frequency link proposed here.

## 2. Combined flywheel oscillator

The frequency link from the $^{87}$Sr transition frequency to the SI second via TAI is concisely expressed in following equation

$$\frac{\nu_0(\text{Sr})}{f_0(\text{SIs})} = \frac{\nu_0(\text{Sr})}{f(\text{LFO})} \frac{f(\text{LFO})}{f(\text{UTC(k)})} \frac{f(\text{UTC(k)})}{f(\text{TAI})} \frac{f(\text{TAI})}{f_0(\text{SIs})}, \tag{1}$$

where $\nu_0(\text{Sr})$, $f_0(\text{SIs})$, $f(\text{LFO})$, $f(\text{UTC(k)})$, and $f(\text{TAI})$ are the frequencies of the Sr clock transition, the SI second, the LFO, the local realization of Universal Coordinated Time (UTC) at laboratory "k" and the frequency of TAI, respectively. The absolute frequency is derived as the product of four ratios. An HM or Cs fountain standard is utilized as an LFO. It is difficult to measure the four ratios with an identical evaluation interval, requiring careful estimation of the possible difference of the mean frequencies with respect to the intervals. For instance, $\nu_0(\text{Sr})/f(\text{LFO})$ is typically measured in $10^3$–$10^4$ s, at which duration the instability of an HM is optimal. On the other hand, $f(\text{UTC(k)})/f(\text{TAI})$ is available with an evaluation interval of five days or its integer multiple. Thus, we need to evaluate the uncertainty in assimilating the five-day mean of UTC(k) with a mean of over $10^3$ – $10^4$ s of the optical clock operation [11]. Considering these points, the formula is expressed as follows with the specification of the evaluation interval,

$$\frac{\nu_0(\text{Sr})}{f_0(\text{SIs})} = \frac{\nu_0(\text{Sr})}{f(\text{LFO};10^4\text{s})} \frac{f(\text{LFO};5\text{d})}{f(\text{UTC(k)};5\text{d})} \frac{f(\text{UTC(k)};5\text{d})}{f(\text{TAI};5\text{d})} \frac{f(\text{TAI};1\text{m})}{f_0(\text{SIs})}. \tag{2}$$

Here, 5d and 1m denote the evaluation intervals of 5 days and 1 month, respectively. Circular T publishes the frequency of TAI as a one-month mean. Thus, $f(\text{TAI};5\text{d})$ and $f$

(TAI;1m) in Eq. (2) are also not identical, requiring careful estimation of the possible difference between the two mean frequencies of TAI. Adjusting the estimation intervals of TAI to 5 days is a solution so that the link between TAI and the SI second does not suffer from additional uncertainty [13].

Some uncertainties were further reduced in this work. The so-called "dead time uncertainty" of the LFO was slightly reduced by employing a combined oscillator. It is known that the change in frequency in HMs is composed of linear frequency drift and random fluctuations. Hereafter, the former is called deterministic part, whereas the latter is called stochastic part [14]. In this work, one measurement campaign lasts for 10 consecutive days, in which the Sr lattice clock is operated for about $10^4$ s each day. As shown later in this article, we first measured the transition frequency based on an HM, and then the linear regression to the 10-day results leads to the 10-day mean of $v_0$(Sr) / $f$ (LFO;10d). This fitting determines the deterministic part of the LFO's behavior and its uncertainty. In terms of the stochastic part, we evaluated the magnitude of the integrated stochastic phase [15], which depends on the phase noise of the LFO. It is known that the phase noise may be reduced by taking the average of multiple atomic clocks [14].

Figure 1(a) shows the Allan deviations of the two HMs (HMa and HMb hereafter) that we used in this work. For the HMs, the Allan deviation in a long averaging time is determined by the deterministic linear drift. Thus, we characterized the behavior of the stochastic phase fluctuation by the three-sample variance of the frequency, namely Hadamard variance. The overlapping Hadamard deviation between HMa and HMb is shown in Fig. 1(b) as empty squares. It shows a flicker frequency noise floor at $5\times10^{-16}$ from a half day to 10 days. For an averaging time of more than 10 days, the overlapping Hadamard deviations $\sigma_{\text{Ha}}(\tau)$ and $\sigma_{\text{Hb}}(\tau)$ of HMa and HMb, respectively, are calculated against UTC from a half-year record that includes the campaigns of this work. $\sigma_{\text{Ha}}(\tau)$ and $\sigma_{\text{Hb}}(\tau)$ are partly determined by the instability of the UTC-UTC(NICT) link, which is estimated to be the dashed line in Fig. 1(b). The flicker frequency noise floor of UTC is 1.8 parts per $10^{16}$ [16], and the link uncertainty between UTC and UTC(NICT) is assumed here to be 0.21 ns to allow it to be used as a guide to the eye, although it is claimed to be 0.3 ns in Circular T. The data clearly shows that $\sigma_{\text{Ha}}(\tau)$ and $\sigma_{\text{Hb}}(\tau)$ are mostly masked by the UTC-UTC(NICT) link over 5 to 10 days. Those over longer than 10 days, however, indicate the stochastic behavior of the two HMs. It is suggested that the flicker frequency noise floor exists in the low $10^{-16}$ level, and that HMb is noisier than HMa up to 30 days. These characteristics imply that $\sigma_{\text{Ha}}(\tau)$ and $\sigma_{\text{Hb}}(\tau)$ have flicker noise floors of $3\times10^{-16}$ and $5\times10^{-16}$, respectively, for an averaging time from a half day to 20 days.

The Hadamard deviation of the combined flywheel oscillator with its phase $(\phi_\text{a} + \phi_\text{b})/2$ is also shown as red empty circles in Fig. 1(b) against UTC, where $\phi_\text{a}$ and $\phi_\text{b}$ denote the phases of HMa and HMb, respectively. For an averaging time of more than 30 days where the stability is no longer masked by the UTC-UTC(NICT) link stability, the Hadamard variance remains lower by combining the two oscillators, suggesting that employing a combined flywheel oscillator may help to reduce the errors relevant to intermittent operations.

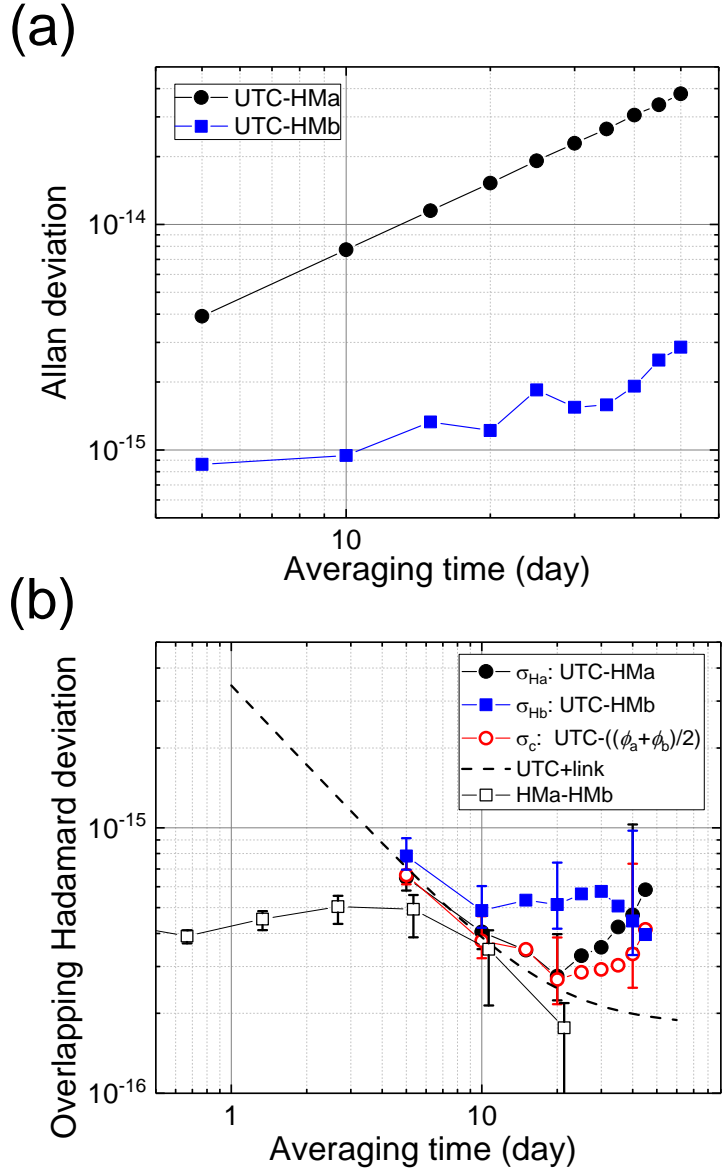

Fig. 1 (a) Allan deviations of the HMs used in this work. (b) Overlapping Hadamard deviations of the two HMs and the combined oscillator. Over the averaging time of up to 10 days, the deviations between the two HMs are shown as empty squares, whereas those over longer times are shown as deviations versus UTC.

Another improvement in this TAI-based measurement is the length of a campaign. One campaign intermittently lasted for 10 days, which is longer than the five days employed in the previous measurement [11], resulting in the reduced uncertainty of the UTC-UTC(NICT) link. The fractional uncertainty is approximately calculated by dividing the link-uncertainty of the time difference by the measurement duration. The time difference between UTC and UTC(NICT) has an uncertainty of 0.3 ns. Thus, the mean frequency ratio $f$ (UTC(k); $n$ day) / $f$ (TAI; $n$ day) has an uncertainty of $(\sqrt{2} \times 0.3 \times 10^{-9})/[(3600 \times 24 \times 5)\left(\frac{n}{5}\right)^{0.9}]$ [9]. A campaign length of 10 days leads to an uncertainty of $4.9 \times 10^{-16}$. Since the link error is random, the three sets of 10-day campaigns performed in this work resulted in a total TAI-UTC(k) uncertainty of $2.9 \times 10^{-16}$.

## 3. Experimental setup

The block diagram of the frequency link is shown in Fig. 2. The setup of the Sr lattice clock is the same as that in the previous work [11] except that the frequency comb has been changed from an ytterbium fiber comb to an erbium fiber comb. Improvements in the evaluation of systematic effects are described in the next section together with the discussion of shifts and uncertainties. The optical signal generated from the lattice clock is down converted to 100 MHz using the frequency comb and microwave frequency divider. The fourth harmonics of the repetition rate (4×250 MHz) is obtained using a high-speed photo detector, and then a ten-fold frequency divider generates a 100 MHz microwave. The phase of this down-converted signal is denoted as $\phi_S$. The phase difference $\phi_S - \phi_a$ is recorded every second by a commercial frequency stability measurement set. By also introducing the same signal to the reference input, we evaluated the system noises of the instrument to be $5\times10^{-15}$ and $3\times10^{-18}$ over averaging times of 1 and $10^4$ s, respectively.

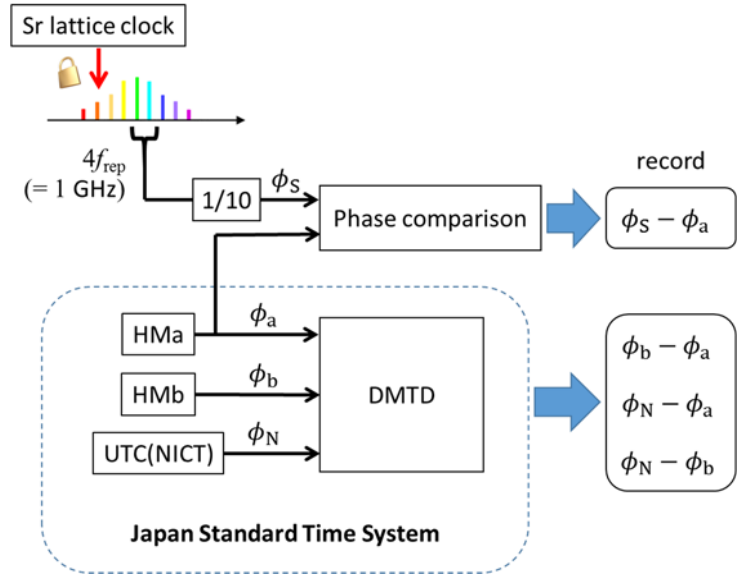

Fig.2 Schematic diagram of the local measurement to obtain $f(\mathrm{Sr})/f(\mathrm{UTC(NICT)};10\mathrm{d})$

We do not need to synthesize a combined signal in order to use the combined flywheel oscillator, but we need to measure the clock frequency based on HMa as well as HMb, and later to obtain the average of two results. Since HMa and HMb are parts of the Japan Standard Time (JST) system, the relative phase (in the unit of second) between the two HMs $\Phi_2 = \phi_b - \phi_a$ is measured in every second by a dual-mixer time difference (DMTD) system [17]. The optical clock was operated for $\Delta t = t_f - t_i \approx 10^4$ s per day. The mean frequency ratio of the two HMs over the duration $\Delta t$ is calculated using the differences between $\Phi_2(t_i)$ and $\Phi_2(t_f)$, resulting in

$$\frac{f(\mathrm{HMb};\Delta t)}{f(\mathrm{HMa};\Delta t)} = 1 + \frac{\Phi_2(t_f) - \Phi_2(t_i)}{t_f - t_i}. \tag{3}$$

Thus, the transition frequency based on HMb is obtained by converting $v_0(\mathrm{Sr})/f(\mathrm{HMa};\Delta t)$ to

$$\frac{v_0(\mathrm{Sr})}{f(\mathrm{HMb};\Delta t)} = \left[1 + \frac{\Phi_2(t_f) - \Phi_2(t_i)}{t_f - t_i}\right]^{-1} \frac{v_0(\mathrm{Sr})}{f(\mathrm{HMa};\Delta t)}. \tag{4}$$

Note that the measurement resolution of the DMTD is 2 ps. The Allan deviation of the system noise floor was measured in [17] to be $1\times10^{-16}$ with an averaging time of $10^4$ s. The JST system also measures the phase difference of HMs against UTC(NICT), which is used to connect the HMs to UTC.

## 4. Systematic shifts and uncertainties of the $^{87}$Sr frequency standard

The systematic shifts and uncertainties of the $^{87}$Sr optical frequency standard are summarized in Table 1. The uncertainties in the blackbody radiation (BBR) shift and the dc Stark shift were reduced over the previous work [11]. Some other effects were evaluated again in the same manner as in the previous work. Only the results of the evaluation in this study are shown in Table 1.

Table 1. Systematic shifts and uncertainties of the $^{87}$Sr lattice clock

| Effect | Shift ($10^{-17}$) | Uncertainty ($10^{-17}$) |
|---|---|---|
| Blackbody radiation | -511.1 | 2.0 |
| Lattice scalar / tensor | -2.7 | 3.8 |
| Lattice hyperpolarizability | 0.2 | 0.11 |
| Lattice E2/M1 | 0 | 0.5 |
| Probe light | -0.1 | 0.1 |
| Dc Stark | -0.12 | 0.16 |
| Quadratic Zeeman | -52.2 | 0.3 |
| Density | -4.1 | 2.8 |
| Background gas collisions | 0 | 1.8 |
| Line pulling | 0 | 0.1 |
| Servo error | 0 | 1.5 |
| Total | -570.1 | 5.7 |

*4.1 Blackbody radiation (BBR)*

The vacuum chamber is made of aluminum alloy of > 20 mm thickness to suppress the spatial inhomogeneity of temperature. The temperature of the lab, $T_{lab}$, is stabilized to be 23.8 ± 0.39 °C during the campaigns. The cooling water of the Zeeman slower tube is supplied from a temperature-controlled circulating system, where the temperature of water is adjusted so that the chamber temperature is same as room temperature within 1 K. Adjusting the chamber temperature to $T_{lab}$ frees us from having to consider the emissivity on the inner surface of the aluminum chamber. Moreover, the physical supports of MOT coils, which are the nearest heat source for atoms, are fixed to the optical table separately from the chamber, preventing conductive heat transfer to the vacuum chamber. The temperature of the vacuum chamber is monitored by three calibrated thermistors with different locations on chamber surface. Owing to these measures to reduce the thermal inhomogeneity, we suppressed the temperature difference among the three sensors to below 0.3 K, although the temporal fluctuation of the mean chamber temperature was slightly larger with a peak-to-peak value of 0.7 K during the experiment. The uncertainty of the calibrated sensors was 0.2 K. For the estimation of the BBR shift, we set the temperature of the aluminum chamber to be the mean temperature of the three thermometers (= $T_{ave}$) with an uncertainty of ($T_{max} - T_{min}$)/2, where $T_{max}$ and $T_{min}$ is the maximum and minimum among the three temperatures, respectively. The vacuum chamber has two glass windows of 45 mm diameter with a distance of 23 mm from the atoms. We assumed that the temperature of the glass window is the midpoint of $T_{ave}$ and $T_{lab}$ with uncertainty of $|T_{ave} - T_{lab}|/2$. In addition, one and six glass windows with 16 and 35 mm diameter are located at a distance of > 120 mm from the atoms. Since the total fractional view angle of these minor windows for atoms is less than 4%, we regarded these minor windows as having the same temperature as the vacuum chamber. When the atoms are interrogated by the clock laser, the atomic beam source is isolated by a mechanical shutter, which is placed about 60 mm away from the nozzle behind a heat shield with a 6-mm aperture. The temperature of the shutter $T_{shutt}$ is estimated to be 371(50) K according to the variation of the resistance of the shutter coil. By considering the solid angle to see the end of Zeeman slower tube from atoms, the contribution of the oven BBR is 7(4) ×10$^{-19}$, having a minor impact to the total uncertainty. By taking into account these issues as well as the BBR coefficient carefully measured in [18] and validated theoretically in [19], the BBR shift is determined to be – 5.111×10$^{-15}$ with an uncertainty of 2.0×10$^{-17}$.

*4.2 Dc Stark shift*

Two glass windows of 45 mm diameter face atoms with a distance of 23 mm. We removed stray static electric charge on these windows by irradiating UV light [20]. The existence of a stray dc electric field was investigated by evaluating the differential measurement of the frequency shift on the polarity of the external electric field of three directions. While the three directions of the external field are not orthogonal, the response indicated null electric charge within the uncertainty. The corresponding fractional uncertainty to the clock frequency was low 10$^{-18}$ level, having no impact on the total uncertainty.

*4.3 Lattice Stark shift*

The polarization of the lattice laser and the bias magnetic field are parallel. We interrogated the π transitions ($m_F = \pm 9/2 \rightarrow m_F = \pm 9/2$). Interleaving evaluations of the ac Stark shift, indicated that the magic frequency of our setup is 368 554 516 (15) MHz, which is 11 MHz smaller than that of previous measurement [11]. The experiment was performed with the lattice depth of 41 $E_R$, where $E_R$ is the recoil energy of a lattice photon. The frequency of the lattice laser was tuned to 368 554 527 MHz. The slight difference and uncertainty resulted in the fractional lattice light shift of -0.27(38) ×10$^{-16}$.

*4.4 Quadratic Zeeman and line pulling*

The center frequency of the clock transition is obtained as the midpoint of the two π transitions from the $m_F = \pm 9/2$ stretched states. The separation of the two transitions indicates the magnitude of the bias magnetic field. The separation was 951.5(1) Hz in the campaigns. This corresponds to the quadratic Zeeman shift of -5.22(03) ×10$^{-16}$, where the shift coefficient was obtained from the weighted mean of five other measurements [21-25]. According to the separation of the two transitions, the transition from the adjacent magnetic sublevel $m_F = \pm 7/2$ is separated by 95 Hz, which is more than eight times larger than the observed spectral width of 10 Hz, suppressing the impact of the line pulling to < 1.4×10$^{-18}$.

*4.5 Servo error*

The linear drift of the stable cavity is pre-compensated using a DDS-based drift-remover. The change in the drift rate is obtained from the result of the spectroscopy in the last 125 s, and the corresponding change is corrected to the drift rate of the DDS. Nevertheless, the existence of a higher-order change causes a systematic error. The average of this error $\bar{e}$ over $T_{free}$ = 125 s is calculated to be $\bar{e} = G^{-1}(1-G)\nu_{cav}T_{free}T_{cycle}$, where $\nu_{cav}$, $G$, and $T_{cycle}$ are the coefficient of the quadrature change of the cavity resonance, the gain, and the cycle time of the digital servo to stabilize the laser frequency to the atomic resonance, respectively. In this work, we adopted $G$ and $T_{cycle}$ of 0.5 and 5 s, and the resultant maximum coefficient $\nu_{cav}$ was observed to be 1.0×10$^{-5}$ Hz/s$^2$. These parameters determined the possible servo error $\bar{e}$ to be 6.5 mHz, which corresponds to a fractional possible error of 1.5×10$^{-17}$.

*4.6 Other systematic shifts*

The optics for the optical lattice path is identical to that in the previous work [11], which was also confirmed by the same magnitude of the coefficient of the lattice Stark shift against the lattice light intensity. We again evaluated the density shift using the alternative atomic servo with the fractional difference in the range between two and ten. The result was consistent with the previous work. Thus, we calculated the coefficient of the density shift with these new samples added. Finally, the shift and the uncertainty of the density shift were concluded to be -0.41(28) ×10$^{-16}$. The lifetime of the atoms trapped in the optical lattice leads to a shift of 0(1.8) ×10$^{-17}$ based on [26].

## 5. Results

We planned three 10-day measurement campaigns on MJD 57474−57484, 57539−57549 and 57584−57594, in which the lattice clock was operated for approximately 10000 s on each day except for one day of no measurement in campaign #3 due to a technical problem with the apparatus. The clock transition was first measured with reference to HMa. The result obtained in campaign #2 is shown in Fig. 3(a). Considering that the dominant phase noise of HMa is white frequency noise up to 1000 s, the measurement was divided into durations of up to 1000 s with error bars determined by the Allan deviation of HMa with the respective averaging time. The linear fitting to these points (100 in total for one campaign) leads to the 10-day mean frequency ratios of $v_0$(Sr) and $f$(HMa). Then, the 100 points were converted to those with reference to HMb using Eq. (4), and the averages of the two fractional frequency ratios

$$\frac{1}{2}\left(\frac{v_0(\text{Sr})}{f(\text{HMa};\Delta t)} + \frac{v_0(\text{Sr})}{f(\text{HMb};\Delta t)}\right) \tag{5}$$

were calculated. The results of the linear fittings to the data based on HMa, HMb, and the combined oscillator are shown in Fig. 3(b). Here, 10 sets of 1000-s data obtained in one day were averaged and expressed as one point. The offsets, in other words, the 10-day means of the fractional difference, were removed so that the three lines intersect at zero. The benefit of employing the combined oscillator was found to be the reduced scattering of the points in the combined oscillator. HMa is more stable than HMb as shown in Fig. 1(b). Thus, the fitting error in the combined oscillator is not always better than that using only HMa. Nevertheless, the square roots of the mean squares of the fitting errors in three campaigns are 1.75, 2.23 and 1.67 parts per 10$^{16}$ for the LFOs of HMa, HMb, and the combined oscillator, respectively. This indicates that combining a more or less stable HM does not strongly deteriorate the stochastic behavior. Combining multiple LFOs, on the other hand, certainly mitigates the problem of possible sporadic excursion in the frequency of one LFO.

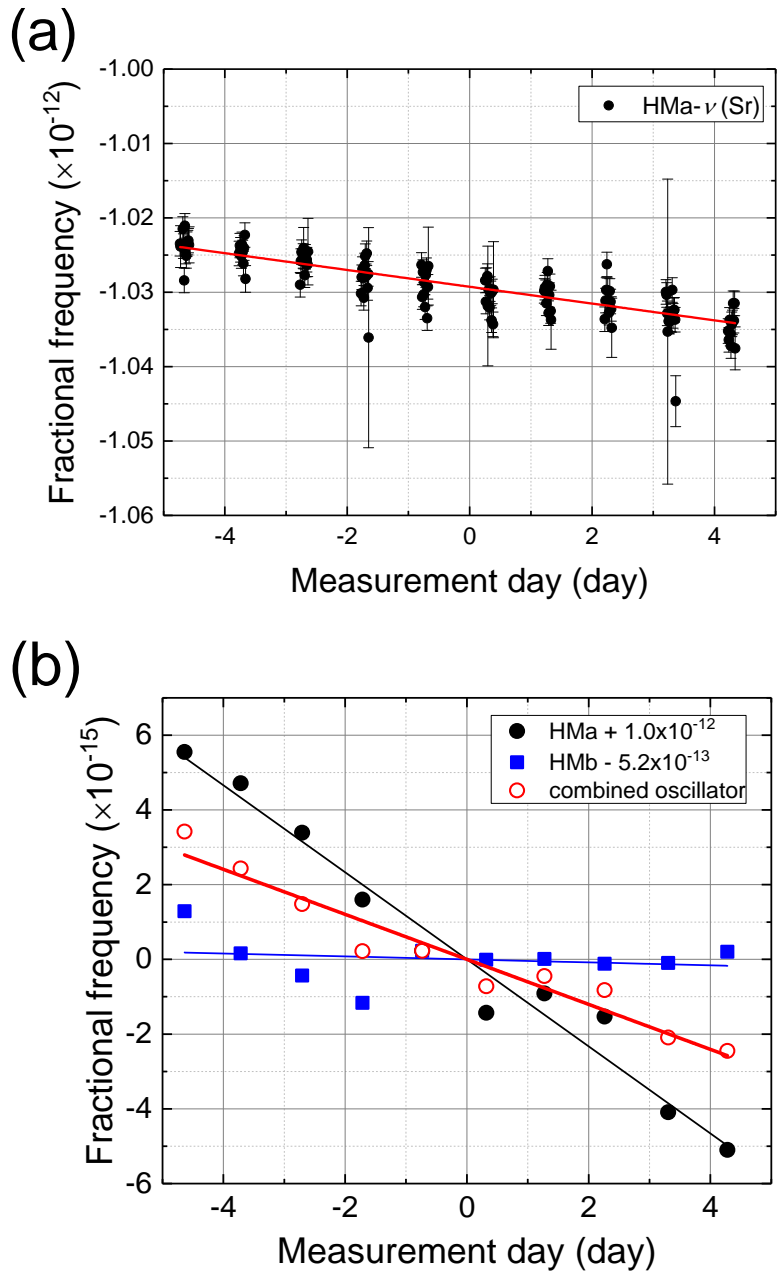

Fig.3 (a) Evaluation of HMa frequency based on Sr in a 10-day intermittent operation (Campaign #2, MJD 57539−57549). Most of the points are the results of $10^3$ s evaluations. The error bars are the Allan deviation of the HMa in the corresponding evaluation interval. Linear fitting provides mean frequency of 10 days. (b) Fractional frequency of the optical clock signal against two HMs and a combined oscillator. The linear frequency drift of the HMs is clearly observed. The offset is removed so that the three linear fitting lines intersect at the midpoint of the campaign. The fluctuation of the data points is mitigated when we take the average of the two values, which is a benefit of employing the combined LFO.

Table 2. Uncertainties of the absolute frequency measurement

|  | Campaign #2 ($10^{-17}$) | Total ($10^{-17}$) |
|---|---|---|
| **Strontium** | | |
| statistical | 2 | 1 |
| systematic | 6 | 6 |
| **Gravity** | 2 | 2 |
| **Local flywheel oscillator** | | |
| deterministic | 18 | 10 |
| stochastic (dead time) | 10 | 6 |
| **Link** | | |
| UTC-UTC(NICT) link | 49 | 28 |
| **UTC- SI second** | (50) | |
| systematic uncertainty | 15 | 14 |
| rest of random part | 48 | 26 |
| Total | 73 | 43 |

Table 2 shows the summary of the link uncertainties to determine the absolute frequency. The uncertainty of the first factor in Eq. (1), namely $\Delta[\nu_0(\text{Sr})/f(\text{LFO};10\text{d})]$, is composed of the statistical and systematic effects of the Sr system, and the uncertainty attributed to the phase fluctuation of the LFO. The short-term instability of our clock laser is at the low $10^{-15}$ level, predominantly limited by the thermal noise of a conventional 10-cm ULE cavity. The alternative operation of two independent servo loops shows an instability of $7\times10^{-15}/\tau^{1/2}$ per servo loop, where $\tau$ is the averaging time in seconds. Thus, the statistical uncertainty of the Sr system in one campaign of more than $10^5$ s of operation is estimated to be $2.2\times10^{-17}$. The systematic shifts and uncertainties of the strontium system were discussed in the previous section. For the measurement with reference to TAI, we need to evaluate the environment of atoms related to gravity. The uncertainty of the gravitational red shift is composed of the static part attributed to the elevation as well as a dynamic part attributed to the tidal effect. The elevation from the geoid surface was evaluated to be 76.5(2) m using the geoid model of Japan, GSIGEO2011 [27]. While the geometrical height was evaluated before the large earthquake in 2011, the change due to the earthquake was less than 5 cm, which we can ignore at this stage [28]. The measurement performed here is affected not by half-day oscillations of the tidal effect but by the 10-day mean of the tidal effect. The residual offset caused by the 10-day mean of the tidal effect is less than 15 cm, which corresponds to $2\times10^{-17}$.

The standard error of the linear fitting as shown in Fig. 3(a), corresponds to the uncertainty of the deterministic part of the combined LFO. The stochastic part is determined by the random fluctuation of the HM phase, which we estimate on the basis of Hadamard deviation. According to Fig. 1(b), the dominant properties of the phase noise are flicker frequency noise and random walk flicker noise, $\sigma_H((\phi_A + \phi_B)/2; \tau) = 3\times10^{-16} + 1.5\times10^{-19}\tau^{1/2}$, where $\tau$ is the averaging time in seconds. The induced random phase $x(\tau_D)$ in the dead time of $\tau_D$ is known to be $x(\tau_D) = \tau_D \sigma(\tau_D)/(\ln 2)^{1/2}$ in the case of flicker frequency noise [15]. Thus, in the one non-operational time of $\tau_D = (86400-10000)$ s, the induced phase noise is $76400\times3\times10^{-16}/(\ln 2)^{1/2} = 27.5$ ps. The random walk frequency modulation is $76400^{3/2}\times1.5\times10^{-19} = 3.2$ ps. Accounting for all the dead times, the rms value of the extra phase in one campaign is estimated to be 85.3 ps. Therefore, the uncertainty in the frequency is $85.3\times10^{-12}/864000 = 9.9\times10^{-17}$. This uncertainty was also estimated by a numerical simulation in the same manner as employed in [10,11]. First, the commercial software program "Stable 32" generated a time series of frequency for $86400 \times 10$ s. Then, the difference in the mean frequency from the total mean over 10000 s × 10 days was calculated. The result was consistent with the uncertainty based on the above simple estimation. The link uncertainty between UTC and UTC(NICT) is 0.3 ns. This yields a fractional uncertainty of $4.9\times10^{-16}$ in the difference of the 10-day mean frequencies of UTC and UTC(NICT) [9].

The ratio between $f$(TAI) and the SI second is derived with respect to the 10 days of the measurement campaigns [13] instead of the one-month mean of $f$(TAI; 1m) / $f_0$(SIs) with an uncertainty of a few parts per $10^{16}$ routinely reported in Circular T. Since the estimation interval is a free parameter in the estimation protocol of the TAI scale [29], it is possible to derive $f$(TAI) / $f_0$(SIs) for the 10-day mean. The uncertainty of $f$(TAI;10d) / $f_0$(SIs) is typically $5\times10^{-16}$, which is larger than that of the conventional one-month mean as most evaluations using PFSs are reported based on the measurement over one-month in order to reduce the link uncertainty. While the shorter estimation interval of 10-day increases the uncertainty of $f$(TAI) / $f_0$(SIs), using the same TAI estimation intervals as the campaign length frees us from having to consider the possible error in assimilating the 10-day mean of TAI to that of one-month mean. Note that the estimation of this possible error requires the rigorous evaluation of the phase noise in TAI, which has not yet been well characterized since the improvement of the algorithm to synthesize EAL [8]. The sources of the uncertainty in $f$(TAI;10d) / $f_0$(SIs) are the systematic uncertainties of the PFS and random uncertainties such as the type-A and link uncertainties as reported in Circular T. According to the method described in [13], the systematic part is $1.5\times10^{-16}$ thanks to the averaging of multiple PFSs. We conservatively assumed that this systematic part is not statistically averaged in the total average of the three campaigns as the majority of the PFSs were common in the three campaigns. On the other hand, the rest of the random part ($4.8\times10^{-16}$) is averaged down by $3^{1/2}$ to obtain $2.7\times10^{-16}$ in the total mean.

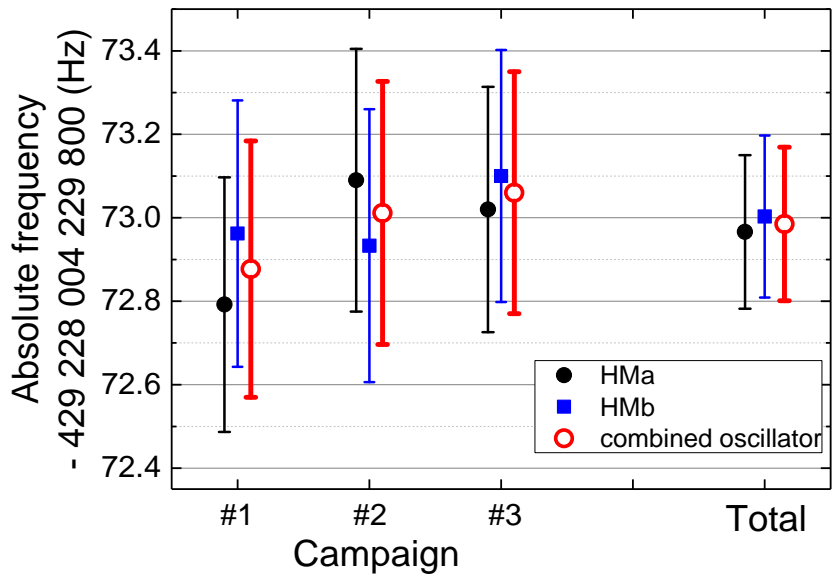

Fig. 4 Absolute frequency of the $^{87}$Sr clock transition evaluated with the LFO of two HMs and the combined oscillator.

The results of the three campaigns and their total mean are shown in Fig. 4, where two single HMs or one combined oscillator was employed as LFOs. The total uncertainty of one campaign with the combined oscillator is $7.3\times10^{-16}$, and it is statistically reduced to $4.3\times10^{-16}$ in the total mean of the three campaigns. Finally, the absolute frequency as the total mean of the three campaigns using the combined LFO was determined to be 429 228 004 229 872.99 (18) Hz. This frequency agrees with other measurements. Figure 5 shows the result obtained in other laboratories, where the top six points were additionally taken into consideration for the determination of the recommended standard frequency of the CIPM 2015 (Comite international des poids et mesures 2015), and the other three points were reported later from PTB and SYRTE with uncertainties below $3\times10^{-16}$, where a large part is attributed to the systematic uncertainty of the local Cs fountains. The frequency reported here using the TAI link is consistent with these highly accurate measurements, suggesting the validity of the measurement scheme investigated here.

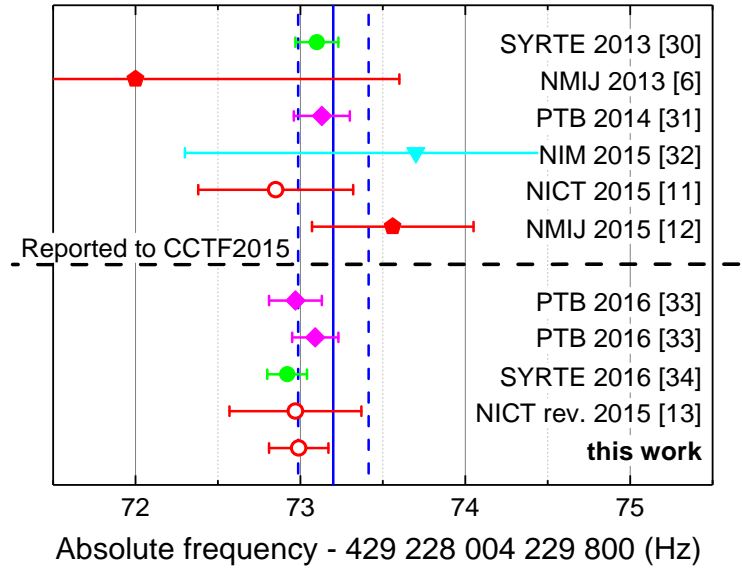

Fig. 5 Comparison of $^{87}$Sr clock transition frequencies recently reported by various institutes [6,11-13,30-34]. The last five points reported in 2016 or later show good agreement, indicating the validity of reducing the uncertainty in the $^{87}$Sr standard frequency.

## 6. Conclusion

The uncertainty of optical frequency measurements in the frequency link between the $^{87}$Sr lattice clock and TAI was reduced to $3.0 \times 10^{-16}$ by employing a combined LFO as well as extending the evaluation interval. The total uncertainty of the absolute frequency is $4.3 \times 10^{-16}$, less than a factor of two larger than those of the most accurate measurements [33,34]. The link technique developed here, on the other hand, enables the evaluation of the TAI frequency with reference to the $^{87}$Sr lattice clock. While the CIPM frequency of the $^{87}$Sr lattice clock still has a fractional uncertainty of $5\times10^{-16}$, four recent results, two in [33], one in [34], and the other in this work, have a fractional standard deviation of $1.2 \times 10^{-16}$. In addition, the reduced χ square value of the four results is 0.14, indicating the consistency among the four measurements. These characteristics suggest that the resultant mean frequency derived from all past measurements is sufficiently reliable to utilize it for other applications such as time scales [33,35] or the calibration of the TAI scale interval.

**Acknowledgements**

It is worth noting that the measurement depends on the Cs fountains at NMIs provided for the calibration of the TAI scale. The authors thank Y. Kobayashi, M. Endo, T. Nakamura, P. Liu, and N. Ohtsubo for their help in setting up the Er fiber system. The discussions with T. Gotoh, M. Kumagai, H. Ito, and M. Hosokawa were greatly appreciated. The stable operation was achieved with the tremendous technical help provided by H.

Ishijima and S. Ito. This work was partly supported by JSPS and MAEDI under Japan - France Integrated Action Program (SAKURA).